\documentstyle[amssymb,preprint,aps]{revtex}

\draft

\begin{document}
\title{Magnetic properties of $a$-Si films doped with rare-earth elements}
\author{M. S. Sercheli and C. Rettori}
\address{Instituto de F\'{i}sica ``Gleb Wataghin'', UNICAMP, 13083-970, Campinas-SP,\\
Brazil}
\author{A. R. Zanatta}
\address{Instituto de F\'{i}sica de S\~{a}o Carlos, USP, 13560-250, S\~{a}o Carlos-SP,%
\\
Brazil }
\maketitle

\begin{abstract}
Amorphous silicon films doped with Y, La, Gd, Er, and Lu rare-earth elements
($a$-Si:RE) have been prepared by co-sputtering and studied by means of
electron spin resonance (ESR), $dc$-magnetization, ion beam analysis,
optical transmission, and $Raman$ spectroscopy. For comparison the magnetic
properties of $laser$-crystallized and hydrogenated $a$-Si:RE films were
also studied. It was found that the rare-earth species are incorporated in
the $a$-Si:RE films in the RE$^{3+}$ form and that the RE-doping depletes
the neutral dangling bonds (D$^{0}$) density. The reduction of D$^{0}$
density is significantly larger for the magnetic REs (Gd$^{3+}$ and Er$^{3+}$%
) than for the non-magnetic ones (Y$^{3+}$, La$^{3+}$, Lu$^{3+}$). These
results are interpreted in terms of a strong exchange-like interaction, $%
J_{RE-DB}{\bf S}_{RE}{\bf S}_{DB}$, between the spin of the magnetic REs and
that of the D$^{0}$. All our Gd-doped Si films showed basically the same
broad ESR Gd$^{3+}$ resonance ($\Delta H_{pp}\approx 850$ Oe) at $g\approx
2.01$, suggesting the formation of a rather stable RE-Si complex in these
films.

75.70.-i, 76.30.Kg, 78.66.Jg
\end{abstract}

\section{Introduction}

The outermost electronic configuration of the rare-earth (RE) elements plays
a decisive role in determining their properties. These elements are known to
be highly electropositive and to exist, predominatly, in the trivalent RE$%
^{3+}$ form.\cite{Cotton} Moreover, RE$^{3+}$ ions have inner $4f^{n}$
electrons that are efficiently shielded by the outermost, and completely
filled, $5s^{2}$ and $5p^{6}$ shells. As a consequence, electronic
transitions involving the $4f$ orbital usually give rise to narrow and well
defined signals that are not, or only weakly, influenced by the local RE$%
^{3+}$ environment. Anticipating the advantages of these attributes, much of
the current interest in studying RE-doped silicon-based compounds arises
from their potential to combine some of the unique characteristics of RE$%
^{3+}$ ions with the electrical properties of semiconductor hosts.\cite
{Polman} Presently, and despite the great advances achieved in the field of
RE-doped semiconductor compounds, the topic is still open with lots of
challenging questions.\cite{Castilho} \cite{Hellman} Specifically related to
the magnetic properties of these compounds, studies on Er-implanted
crystalline silicon have concluded that the ESR signal associated with the Er%
$^{3+}$ ions is absent in samples with no oxygen, which is believed to
stabilize sites for the ions.\cite{Carey} Also, the co-doping with oxygen
(or other light impurities) considerably improves the Er-related
luminescence signal.\cite{Priolo} Similar research have been conducted on
Er-doped hydrogenated amorphous Si-O films and indicated that the
photoluminescence intensity depends on the density of neutral Si dangling
bonds (D$^{0}$).\cite{Konstantinova}

Based on the above scenario, this work presents a systematic study on the
magnetic properties of amorphous silicon ($a$-Si) films doped with different
RE (Y, La, Gd, Er, and Lu) elements. The samples were prepared by the
co-sputtering technique because of its versatility in producing films with
quite different, and controllable, atomic compositions. For comparison
hydrogenated $a$-Si films doped with RE were also analyzed. Complementary
spectroscopic techniques and $laser$-induced crystallized films were also
used in order to achieve further insight.

\section{Experimental}

This work sumarizes and presents part of the data taken from more than 60
films. All films were prepared in a high vacuum chamber (base pressure $\sim
2$ x10$^{-6}$ Torr), by radio frequency ($13.56$ MHz) sputtering a Si ($%
99.999$ \%) target covered at random with small pieces of metallic RE ($99.9$
\%) elements. Polished crystalline ($c$-)Si wafers and high-purity quartz
plates were used as substrates in every deposition run. During deposition,
the substrates were kept at $\sim 70$ 
${{}^o}$%
C under a constant total pressure of $\sim 5$ x10$^{-3}$ Torr consisting of
high-purity gases (Ar or a mixture of Ar + H$_{2}$, for the hydrogenated
films). For the whole series of films the RE concentration was determined by
the relative RE-to-Si target area (A$_{\text{RE}}$/A$_{\text{Si}}$).
Non-hydrogenated $a$-Si and hydrogenated $a$-Si ($a$-Si:H) films were also
deposited for comparison purposes. $Laser$-induced crystallization
treatments, at room atmosphere, were done on some of the RE-doped $a$-Si and 
$a$-Si:H films deposited on quartz substrates. For this treatment,
cylindrical lenses and the $532.0$ nm line of a Nd-YAG $laser$ (pulse
duration of $10$ ns, and repetition of $5$ Hz) were employed rendering a $%
laser$ fluence of $\sim 500$ mJ cm$^{-2}$. \cite{Bell}

The atomic composition of the films were determined mostly from $Rutherford$
backscattering spectrometry (RBS) and nuclear reaction analysis (NRA). The
optical band gap of the films were investigated through optical transmission
in the visible-ultraviolet range in a commercial spectrophotometer. $Raman$
scattering measurements, at room-$T$ and with the $514.5$ nm line of an Ar$%
^{+}$ $laser$, were also performed to analyze the atomic structure of the
RE-doped $a$-Si(:H) films.

The electron spin resonance (ESR) experiments were carried out in the $300-4$
K $T$-range in\ a $Bruker$ X-band ($9.47$ GHz) spectrometer using a room-$T$
TE$_{102}$ cavity. The $dc$-magnetization measurements were accomplished in
the $300-2$ K $T$-range using\ a $Quantum$ $Design$ SQUID magnetometer (RSO
mode, $1$ T). In addition to the RBS data, the concentration of the magnetic
RE species was also determined from the ESR and $dc$-susceptibility data, $%
\chi $($T$). In the former, [RE] has been obtained after comparison with a Gd%
$_{1.45}$Ce$_{0.55}$RuSr$_{2}$Cu$_{2}$O$_{10+\delta }$ and a strong
KCl-pitch standard samples and in the latter, following a fitting of the low-%
$T$ $dc$-susceptibility data to a {\it Curie-Weiss} law, $\chi (T)=ng^{2}\mu
_{B}^{2}J(J+1)/3k(T-\theta _{p})$, after subtraction the diamagnetism of a
similar undoped thin film/substrate system.

\section{Results and Discussion}

Table I displays P(H$_{2}$), A$_{\text{RE}}$/A$_{\text{Si}}$, and the atomic
concentration [RE] (as determined from RBS, NRA, ESR, and $\chi $($T$) data)
of some of the films investigated in this work. For simplicity $laser$%
-crystallized silicon ($lc$-Si) films are not included in this Table. Notice
that in spite of the quite different methods employed to determine [RE] we
always found [RE] $\approx $ A$_{\text{RE}}$/A$_{\text{Si}}$. Nevertheless,
considerable deviations occur for films deposited in a Ar + H$_{2}$
atmosphere. This is an expected result since we have adopted a constant
total pressure of $\sim 5$ x10$^{-3}$ Torr in all deposition runs. Thus,
taking into account the different sputter yield due to H$_{n}^{+}$ and Ar$%
^{+}$ ions,\cite{Chapman} the concentration of REs in the hydrogenated films
is expected to decrease as P(H$_{2}$) increases. Table I shows that, within
the experimental error, the magnetic RE contents estimated from ESR and $%
\chi $($T$) measurements agree with those obtained from RBS, indicating that
most of the RE species are indeed incorporated into the $a$-Si(:H) films in
the RE$^{3+}$\ form. For the Gd$^{3+}$ and Er$^{3+}$-doped $a$-Si(:H) films
the $\chi $($T$) data follow a $Curie-Weiss$ law at low-$T$ (not shown). The
fitting of the $\chi $($T$) data to a $Curie-Weiss$ law shows that the
obtained paramagnetic temperature, $\theta _{p}$, of the Gd-doped $a$-Si(:H)%
{\em \ }films is negative and larger for films of higher Gd concentration,
indicating the existence of an AFM exchange-like interaction between the Gd$%
^{3+}$ spins (see Table I). \cite{Taylor} The same behaviour was observed in
the Gd$^{3+}$-doped $lc$-Si films and, also, $\theta _{p}$ was larger for
the $lc$-Si than for the corresponding pure $a$-Si film (not shown).

Figure $1$ shows $\chi $($T$) for the undoped films studied in this work.
According to these data both the $a$-Si and the $lc$-Si exhibit a low{\em -}$%
T${\em \ }paramagnetic behaviour, while the $a$-Si:H film is diamagnetic.
This is understood based on the nature of the defects present in these
films: $i)$ non-hydrogenated $a$-Si and $lc$-Si films are known to have,
respectively $\sim 10^{20}$ and $\sim 10^{18}$ spins/cm$^{3}$, a relatively
high density of paramagnetic centers due to singly occupied dangling bond
states, D$^{0}$, \cite{Street1} and $ii)$ hydrogenated $a$-Si films, on the
contrary, exhibit a high density of diamagnetic dangling bonds (either D$%
^{+} $ or D$^{-}$ states). This interpretation is consistent with the
transport properties of the $a$-Si and $a$-Si:H films.\cite{Street1} The
difference in the density of paramagnetic states is evident in the inset of
Fig. 1 that shows the ESR signal of the films being considered. The ESR
signal of a quartz substrate is also shown for comparison.

The ESR line width of the D$^{0}$ signal ($\Delta H_{pp}$) in the $a$-Si and 
$lc$-Si films was measured as a function of $T$ and the results are
displayed in Figure $2$. As compared to the $lc$-Si films, $\Delta H_{pp}$\
in the $a$-Si film presents larger residual width and larger thermal
broadening. The larger residual width\ may be due to an inhomogeneous
broadening and the larger thermal broadening to a stronger spin-lattice
coupling (shorter spin-lattice relaxation time, $T_{1}$), and both caused by
the larger disorder in the $a$-Si films. At the lowest temperature, a small
but detectable line broadening is observed in the $lc$-Si film which is
probably associated with short-range magnetic correlations. The intensity of
these signals increases at low-$T$ and follows approximately a $T^{-1}$
behaviour which is the typical behaviour of localized spins (see inset of
Figure $2$). Within the accuracy of our measurements, the $g$-values are $T$%
-independent.

Figure $3$ shows the D$^{0}$ ESR signal of non-hydrogenated $a$-Si films
doped with magnetic (Gd$^{3+}$ and Er$^{3+}$) and non-magnetic (Y$^{3+}$, La$%
^{3+}$, and Lu$^{3+}$) RE elements ([RE] $\sim 2.5$ at.\%). The ESR signal
of a quartz substrate, $a$-Si, $lc$-Si, and $a$-Si:H films are also shown
for comparison. According to these data RE-doping reduces the ESR signal
intensity of D$^{0}$ states and the reduction produced by Gd$^{3+}$ and Er$%
^{3+}$ is remarkably greater than that caused by Y$^{3+}$, La$^{3+}$, and Lu$%
^{3+}$.

At this point, we shall compare the density reduction of D$^{0}$ states in
our $a$-Si films due to the different REs. The density of D$^{0}$ states and
the $g$-value of all the RE-doped $a$-Si and $lc$-Si films considered in
this work are shown in Figure $4$. For the present discussion it is
important to mention that [RE] $\sim 2.5$\ at.\%, the $laser$%
-crystallization treatment, and P[H$_{2}$] were all similar. Strikingly, and
according to Figure $4$(a) we observe: $i)$ a depletion of D$^{0}$ states
with RE-doping. This is an unexpected result since the doping of $a$%
-semiconductors is well known to increase the density of defects,\cite
{Street2} and $ii)$ among all studied RE ions, Gd$^{3+}$ and Er$^{3+}$ are
the most efficient ones to deplete the density of D$^{0}$ states. These
results suggest that this phenomenon is partly caused by the size and/or
coordination of the RE ions, as exemplified by the density drop of D$^{0}$
states by the non-magnetic Y$^{3+}$, La$^{3+}$, and Lu$^{3+}$ ions (see
Figure 3). It is interesting to observe that, as far as the suppresion of
the D$^{0}$\ states is concerned, the doping with non-magnetic RE ions
induces almost the same effect than the $laser$-crystallization of the $a$%
-Si films (see Figure 3). Moreover, the depletion effect is even more
pronounced for the magnetic RE ions, Gd$^{3+}$\ and Er$^{3+}$, being
stronger for Gd$^{3+}$. These results suggest that this $"extra"$\ depletion
efficiency should be related to the spin of the magnetic REs and that an
exchange-like coupling, $J_{RE-DB}{\bf S}_{RE}{\bf S}_{DB},$ between the RE$%
^{3+}$ spin, $S_{RE}$, and the spin of the D$^{0}$ states, $S_{DB}$, may be
the responsible mechanism. We argue that such a coupling may shift and
broader the D$^{0}$ ESR line beyond our signal detection limits. This
coupling is usually governed by the de Gennes factor $[(g_{J}-1)^{2}J(J+1)]$
or by the spin factor $S(S+1)$. The importance of these factors have been
already adviced in RE-doped superconductor compounds through the supression
of $T_{c}$ caused by the magnetic RE ions. \cite{Maple} \cite{Gschneider}
Notice that the de Gennes and spin factors assume their highest value at the
Gd$^{3+}$ ion ($J=S=7/2$), which is in agreement with the results of Figure $%
4$(a). These results are also consistent with our recent report on the
depletion of neutral dangling bonds, D$^{0}$, by RE doping in a-SiN films. 
\cite{Mauricio}

The analysis of the $g$-values displayed in Figure $4$(b) leads to
interesting conclusions. The $g$-values of the $a$-SiRE films are $\sim
2.0048$ and those for the $lc$-SiRE films $\sim $ $2.0031$, which are close
to the $g$-value of singly charged Si neutral dangling bonds, $\sim 2.0055$
(D$^{0}$ states).\cite{Street1} However, the $a$-SiEr and $lc$-SiEr films
present $g$-values of $\sim 2.0030$ and $\sim 2.0005$, respectively, which
are significantly smaller than those of the corresponding series. These
small $g$-values seem to be intrinsically associated with the Er-doping
process itself and suggests that $n$-type films $(g\longrightarrow 1.998)$ 
\cite{Finger} are obtained as a result of the insertion{\em \ }of Er$^{3+}$
ions. This is in agreement with the results achieved by other authors. \cite
{Priolo}\cite{Konstantinova} \cite{Finger}{\em \ }The exceptional ability of
Er$^{3+}$\ to produce $n$-type films may suggest that the Er$^{3+}$\ ground
state multiplet ($J=15/2$) lies close to the film%
\'{}%
s conduction band. The same trend was also observed for the series of $a$%
-SiRE:H films (not shown). However, due to the weak D$^{0}$ ESR signal
presented by these films{\em \ }(see Figure 3) it was not possible to
accurately determine the $g$-values in these films.

Figure $5$ shows a series of $a$-Si films doped with Gd in the $\sim 0-7$
at.\% range, as determined from RBS (see Table I). This series was
exhaustively studied and for [Gd] $\gtrsim 0.5$ at.\% the spectra show a
broad ($\Delta H_{pp}\approx 850$ Oe) and single resonance ($g\approx 2.01$)
associated to a powder-like spectrum of Gd$^{3+}$. Again, the strong
reduction effect that the Gd$^{3+}$ ions exert on the D$^{0}$ ESR intensity
can be appreciated in Figure $5$. Table I presents the data for a serie of $%
a $-SiGd:H films where it is observed that the Gd incorporated in the films
decreases as [H] increases. Surprisingly, the $g$-value and line width, $%
\Delta H_{pp}$, were independent of the Gd$^{3+}$\ concentration, chemical
environment (amorphous or $laser$-crystallized), H content, and temperature
for $T\gtrsim 30$\ K.\ Hence, these results demonstrate unambiguously that
the Gd$^{3+}$ ions form a very stable complex with Si. Films of relatively
higher Gd concentrations $(\gtrsim 4$ at.\%$)$ present a line broadening for 
$T\lesssim $\ $30$\ K, indicating the existence of Gd$^{3+}$- Gd$^{3+}$
magnetic correlations, in agreement with the large $\mid \theta _{p}\mid $
measured for these films (see Table I). However, in contrast to the work of
Hellman $et$\ $al.$\cite{Hellman}, low field $0.1\leqslant H\leqslant 0.5$\
kOe zero field and field cooled susceptibility measurements in these films
did not show any measurable spin glass-like behaviour down to $T\approx 2$ K.

As stated above, hydrogen atoms in $a$-Si are known to considerably reduce
the density of D$^{0}$ states as a result of the passivation of Si dangling
bonds.\cite{Street2} Actually, most of the devices based on $a$%
-semiconductors correspond to hydrogenated compounds.\cite{Finger} In
addition to the passivation of dangling bonds, the insertion of hydrogen in $%
a$-Si materials promotes the widening of the optical band gap as a
consequence of the recession of the top of the valence band due to the
replacement of Si-Si by Si-H bonds. Therefore, the study of the combined
effects due to the insertion of RE ions and hydrogen atoms in the $a$-Si
host is of outmost importance to our discussion. With this purpose we have
investigated a series of Gd-doped $a$-Si:H films. Figure $6$ shows the
transmission spectra of some Gd-doped $a$-Si films deposited with increasing
hydrogen partial pressures (Table I). According to this figure, films
deposited at higher P(H$_{2}$) exhibit a transmission cutoff at higher
energies, an indication that the optical band gap is indeed being increased.
That is shown in the inset of Figure $6$ where the E$_{04}$ optical band gap
(energy corresponding to an absorption coefficient of 10$^{4}$ cm$^{-1}$) is
represented as a function of the hydrogen content. The E$_{04}$ optical band
gap of other films investigated in the present study are also displayed for
comparison. Usually, films deposited with no hydrogen, and after $laser$%
-crystallization, present an E$_{04}$ optical band gap lower than $\sim 1.5$
eV. Increasing contents of hydrogen widen the optical band gap of these
films which is proportional to [H]. The doping of $a$-Si(:H) films with REs
has the opposite effect with E$_{04}$ values ranging from $\sim 1.1$ to $0.7$
eV, depending on the [RE]. $Laser$-crystallization experiments also reduce
the optical band gap of the RE-doped $a$-Si:H films essentially because of
the removal of Si-H bonds.\cite{Bell}

In addition to hydrogenation, different atomic environments may affect the
magnetic properties of the RE-doped $a$-Si films. Thus, it is important to
characterize the atomic structure of these films. By means of $Raman$
scattering we have studied the\ $laser$-induced crystallized films and the
main results are displayed in Figure $7$. For simplicity, just the spectra
of {\it c}-Si and hydrogenated (and Gd-doped) $a(lc)$-Si films are
presented. This figure shows that the spectrum of $a$-Si:H consists of a
weak and broad $Raman$ signal at $\sim 480$ cm$^{-1}$ that corresponds to
the transverse-optical TO-like mode of a highly distorted $a$-Si network. 
\cite{Lannin} Moreover, non-hydrogenated $a$-Si (either RE-doped or not)
films exhibit (not shown) an extremely faint and almost featureless signal
characteristic of amorphous quasi-metallic compounds. The $laser$%
-crystallization processing of the $a$-SiGd(:H) films gives rise to a
relatively strong scattering contribution at $\sim 515$ cm$^{-1}$ indicating
the presence of ordered Si-Si bonds. Based on the position, and line width,
of this $Raman$ scattering signal it was possible to infer the size of these
so-called crystallites (or crystalline grains) which are $\sim 50$ \AA\
large.\cite{Iqbal}\ Notice that, despite the appearance of these
crystallites, the films still present some amorphous contribution at $480$ cm%
$^{-1}$ indicating that they were not completely crystallized. The
crystallized material was estimated\cite{Okada} from the ratio between the
areas under the $Raman$ peaks due to crystallites and the amorphous tissue, $%
\rho _{c}$/$\rho _{a}$. The inset of Figure $7$ shows $\rho _{c}$/$\rho _{a}$
for various of our studied films. Interestingly, the crystalline fraction $%
\rho _{c}$/$\rho _{a}$ in the $laser$-crystallized films depends strongly on
the presence of RE species. According to Figure $7$, the $lc$-SiGd:H films
exhibit an amorphous contribution substantially smaller than the $lc$-Si:H
films, suggesting that the RE species may act as crystallization seeds in
the $a$-Si:H network in a similar manner to that verified in other
metal-containing $a$-Si:H films. \cite{Nast}\cite{Chambouleyron} However,
the crystalization fraction seems to be weaky dependent on the RE and [RE]
(see inset of Figure 7). This result is in agreement with the trends
discused above where the presence of REs diminish the density of D$^{0}$
states in the films.

Finally, it is opportune to notice that in spite of the great influence
exherted by hydrogenation, RE-doping, and $laser$-crystallization treatment
on the optical and structural properties of the studied $a$-Si films, their
magnetic characteristics seem to remain unchanged. One possible reason for
such a behaviour could rely on the formation of very stable RE-Si complexes
which are almost insensitive to the local environment. This phenomenun is
not completely clear at present and more investigations are under way.
Because the [O] in our films were found to be smaller than 0.5 at.\% (RBS),
the formation of \ any stable RE-Si-O complexes seems to be unlikely.

\section{Concluding remarks}

Rare-earth doped amorphous silicon films were prepared by co-sputtering and
investigated by different magnetic and spectroscopic techniques (ESR, $dc$%
-susceptibility, ion beam analyses, optical transmission, and $Raman$
scattering). Hydrogenated and $laser$-crystallized films were also
considered in the present study. The main experimental results are: $i)${\em %
\ }RE ions are incorporated in the $a$-Si(:H) host, predominantly, in the
trivalent form, $ii)$ $a$-Si and $lc$-Si films exhibit D$^{0}$ spin
densities of $\sim 10^{20}$\ cm$^{-3}$\ and $\sim 10^{18}$\ cm$^{-3}$,
respectively, which are strongly reduced by the RE-doping. The magnetic Gd$%
^{3+}$\ and Er$^{3+}$\ ions present an even stronger density depletion of D$%
^{0}$ states, which led us to suggest an exchange-like coupling between the
spin of the magnetic REs$^{3+}$ and the spin of silicon neutral dangling
bonds. We should mention that the same behavior has been consistently
observed in a series of a-SiN alloys doped with Y, La, Pr, Nd, Sm, Gd, Tb,
Dy, Ho, Er, Yb, and Lu, giving further support to the assumption of the
existence of an exchange-like coupling mechanism,\cite{Mauricio} $iii)$\ The
Gd$^{3+}$ ESR line shape, line width, and $g$-value does not change with
[Gd] or [H], $T$, nor with the local environment (if amorphous or partially
crystallized). These results suggested the formation of an extremely stable
Gd-silicon-like complex. At the moment more research is under way to
elucidate this point, and $iv)$ The reduction of the D$^{0}$ $g$-value in
both, $a$-SiEr and $lc$-SiEr films, indicates that the Er-doping is more
efficient in producing $n$-type films.

\section{Acknowledgments}

The authors are indebted to Professor F.C. Marques (UNICAMP) for critical
reading of the manuscript. Professor F.L. Freire Jr. (PUC- RJ) and Professor
F. Ikawa (UNICAMP) are also acknowledge for the ion beam analyses and for
the access to the Raman facilities, respectively. This work has been
supported by FAPESP, CAPES, CNPq, NSF, and DMR.

\begin{figure}[tbp]
\caption{{}$T$-dependence of the $dc$-susceptibility, $\protect\chi (T)$,
for the $a$-Si, $lc$-Si, and $a$-Si:H films. The inset shows the D$^{0}$ ESR
signal at $300$ K and $9.47$ GHz for the same films.}
\label{Figure 1}
\end{figure}
\begin{figure}[tbp]
\caption{$T$-dependence of the D$^{0}$ ESR line width, $\Delta H_{pp}(T)$,
for the $a$-Si and $lc$-Si films. The inset shows the $T$-dependence of the D%
$^{0}$ ESR intensity for the same films. The lines are guide to the eye. }
\label{Figure 2}
\end{figure}
\begin{figure}[tbp]
\caption{D$^{0}$ ESR spectra at $300$ K of $a$-Si films doped with different
REs. The spectra of a $lc$-Si,\ an $a$-Si:H film and a quartz substrate are
also included for comparison. }
\label{Figure 3}
\end{figure}
\begin{figure}[tbp]
\caption{Density of D$^{0}$ states (a) and $g$-value (b) of $a$-Si and $lc$%
-Si films doped with Y, La, Gd, Er, and Lu. The error bars take into account
experimental uncertainties and films with different contents of RE. The
lines are guides to the eye. }
\label{Figure 4}
\end{figure}
\begin{figure}[tbp]
\caption{ESR spectra of Gd$^{3+}$ in $a$-Si films doped with different
concentrations of Gd. The ESR signal from the quartz substrate is also
displayed. }
\label{Figure 5}
\end{figure}
\begin{figure}[tbp]
\caption{Transmission spectra, as a function of the photon energy, of some
Gd-doped $a$-Si:H films deposited under increasing partial pressures of
hydrogen. The fringe pattern in the spectra refer to interference effects
between the film and the quartz substrate. The inset shows the E$_{04}$
optical band gap as a function of the hydrogen content for some of the films
studied in this work. The E$_{04}$ and [H] values typically found for a-SiRE
and $lc$-SiRE:H films are also represented for comparison.}
\label{Figure 6}
\end{figure}
\begin{figure}[tbp]
\caption{Room-temperature $Raman$ scattering spectra for few of our $a$-Si:H
films. All curves have been normalized, and the spectra of a
non-intentionally doped $a$-Si:H film and of a Si 
\mbox{$<$}%
111%
\mbox{$>$}%
wafer are also displayed for comparison. Notice the effect due to the $laser$%
-crystallization processing on the $Raman$ spectrum of the $lc$-Si:H and $lc$%
-SiGd:H films. The scattering signals at $\sim 480$ and $520$ cm$^{-1}$ are
due to distorted Si-Si bonds and to the TO mode of Si crystals. The inset
shows the crystalline fraction $\protect\rho _{c}$/$\protect\rho _{a}$ in
some RE-doped $lc$-Si(:H) films as a function of [RE]. }
\label{Figure 7}
\end{figure}

Table I - Deposition parameters and compositional data of the RE-doped $a$%
-Si(:H) films. Deposition total pressure (Ar or Ar + H$_{2}$) $\approx $ $5$
x10$^{-3}$ Torr. Substrate temperature $\sim 70$ 
${{}^o}$%
C. P(H$_{2}$), A$_{\text{RE}}$/A$_{\text{Si}}$, [H]$_{\text{NRA}}$, and [RE]$%
_{\text{RBS}}$ are the hydrogen partial pressure, relative RE-to-Si target
area, hydrogen content determined from NRA, and RE content determined from
RBS, respectively. [RE]$_{\text{ESR}}$ and [RE]$_{\chi }$ are the RE content
determined from ESR and $\chi $($T$), respectively. [H]$_{\text{NRA}}^{\star
}$ are values estimated from the absorption spectra in the infrared range. $%
\theta _{p}$ is the paramagnetic temperature obtained from a {\it Curie-Weiss%
} law fitting of $\chi $($T$). $n.a.$ means not available.

$
\begin{tabular}{cccccccc}
Sample & P(H$_{2}$) & A$_{\text{RE}}$/A$_{\text{Si}}$ & [H]$_{\text{NRA}}$ & 
[RE]$_{\text{RBS}}$ & [RE]$_{\text{ESR}}$ & [RE]$_{\chi }$ & $\theta _{p}$
\\ 
& (Torr) & (\%) & (at.\%) & (at.\%) & (at.\%) & (at.\%) & (K) \\ 
a-Si & $<2\times 10^{-6}$ & $0$ & $<1$ & $0$ & $0$ & $0$ & $-1.2(4)$ \\ 
a-Si:H & $5\times 10^{-4}$ & $0$ & $\sim 10^{\ast }$ & $0$ & $0$ & $0$ & $%
n.a.$ \\ 
----------- &  &  &  &  &  &  &  \\ 
a-SiY & $<2\times 10^{-6}$ & $\sim 5$ & $<1$ & $2.5$ & $n.a.$ & $n.a.$ & $%
-6(2)$ \\ 
a-SiY:H & $<2\times 10^{-6}$ & $\sim 7$ & $3.5$ & $1.0$ & $n.a.$ & $n.a.$ & $%
-4(1)$ \\ 
----------- &  &  &  &  &  &  &  \\ 
a-SiLa & $<2\times 10^{-6}$ & $\sim 5$ & $<1$ & $n.a.$ & $n.a.$ & $n.a.$ & $%
-2.0(5)$ \\ 
----------- &  &  &  &  &  &  &  \\ 
a-SiGd (1) & $<2\times 10^{-6}$ & $\sim 10$ & $<1$ & $7.5$ & $7.6$ & $10(2)$
& $-12(2)$ \\ 
a-SiGd (2) & $<2\times 10^{-6}$ & $\sim 5$ & $<1$ & $4.0$ & $5.7$ & $6(1)$ & 
$-7(2)$ \\ 
a-SiGd (3) & $<2\times 10^{-6}$ & $\sim 3$ & $<1$ & $2.5$ & $3.8$ & $5(1)$ & 
$-7(2)$ \\ 
a-SiGd (4) & $<2\times 10^{-6}$ & $\sim 2$ & $<1$ & $1.5$ & $1.4$ & $2.0(5)$
& $-4(1)$ \\ 
a-SiGd (5) & $<2\times 10^{-6}$ & $\sim 1$ & $<1$ & $1.0$ & $1.5$ & $1.5(5)$
& $-3(1)$ \\ 
a-SiGd (6) & $<2\times 10^{-6}$ & $\sim 0.6$ & $<1$ & $0.5$ & $1.3$ & $%
0.5(1) $ & $-4(1)$ \\ 
----------- &  &  &  &  &  &  &  \\ 
a-SiGd:H (1) & $<2\times 10^{-6}$ & $\sim 3$ & $1.2$ & $2.5$ & $4.7$ & $5(1)$
& $-9(2)$ \\ 
a-SiGd:H (2) & $3\times 10^{-5}$ & $\sim 3$ & $2.7$ & $2.0$ & $2.7$ & $4(1)$
& $-6(2)$ \\ 
a-SiGd:H (3) & $1\times 10^{-4}$ & $\sim 3$ & $3.7$ & $2.3$ & $1.1$ & $5(1)$
& $-6(2)$ \\ 
a-SiGd:H (4) & $3\times 10^{-4}$ & $\sim 3$ & $7.5$ & $1.5$ & $1.2$ & $%
2.0(5) $ & $-8(2)$ \\ 
a-SiGd:H (5) & $3\times 10^{-3}$ & $\sim 3$ & $9.0$ & $0.4$ & $n.a.$ & $%
0.2(1)$ & $-0.4(1)$ \\ 
----------- &  &  &  &  &  &  &  \\ 
a-SiEr & $<2\times 10^{-6}$ & $\sim 5$ & $<1$ & $2.5$ & $n.a.$ & $2.0(5)$ & $%
-4(1)$ \\ 
a-SiEr:H & $5\times 10^{-4}$ & $\sim 5$ & $\sim 10^{\ast }$ & $0.5$ & $n.a.$
& $1.0(5)$ & $-4(1)$ \\ 
----------- &  &  &  &  &  &  &  \\ 
a-SiLu & $<2\times 10^{-6}$ & $\sim 4$ & $<1$ & $2.5$ & $n.a.$ & $n.a.$ & $%
-5(1)$ \\ 
a-SiLu:H & $5\times 10^{-4}$ & $\sim 7$ & $5.0$ & $0.4$ & $n.a.$ & $n.a.$ & $%
-5(1)$%
\end{tabular}
$

\bigskip

\bigskip

\bigskip

\bigskip

\bigskip

\bigskip

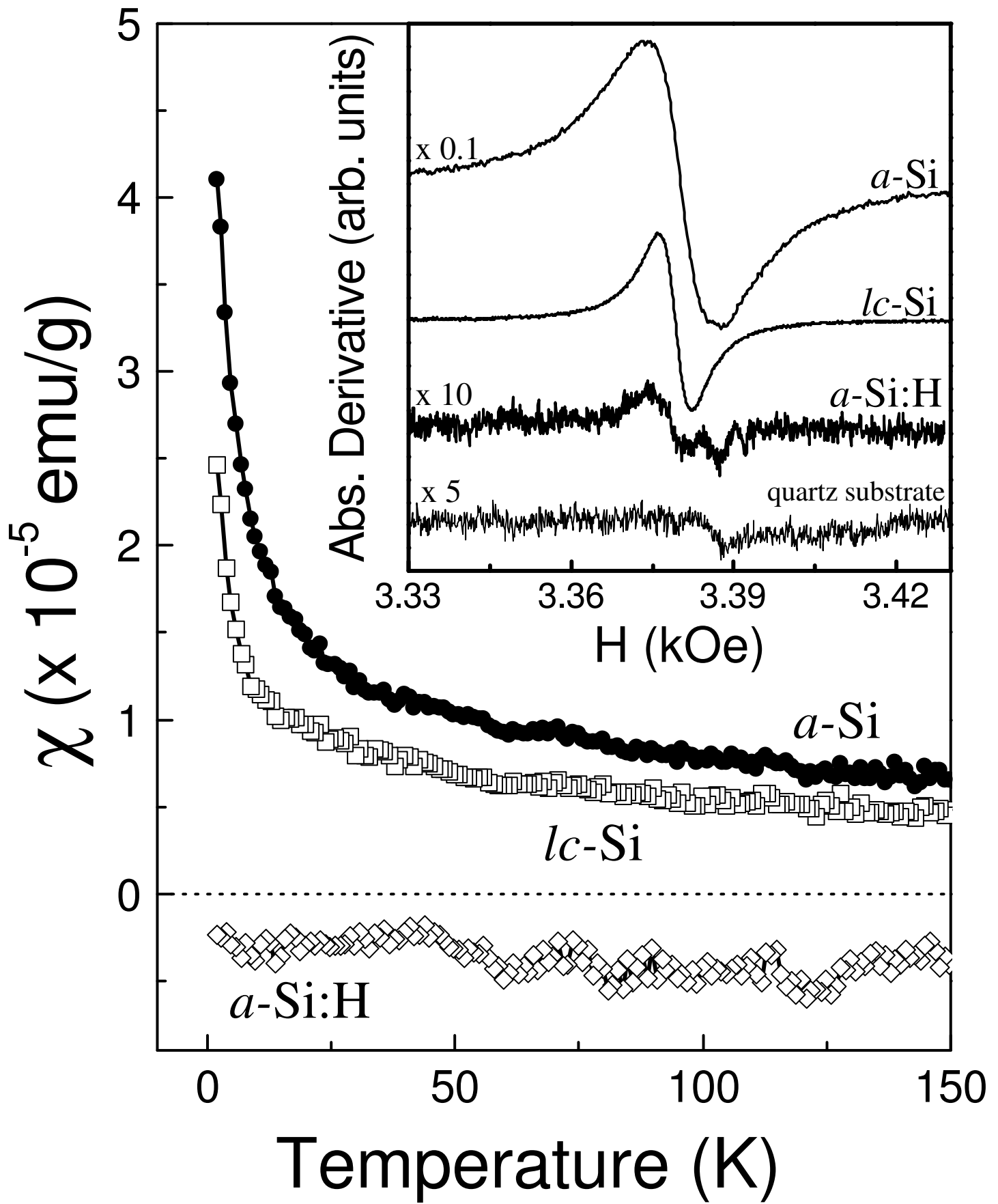
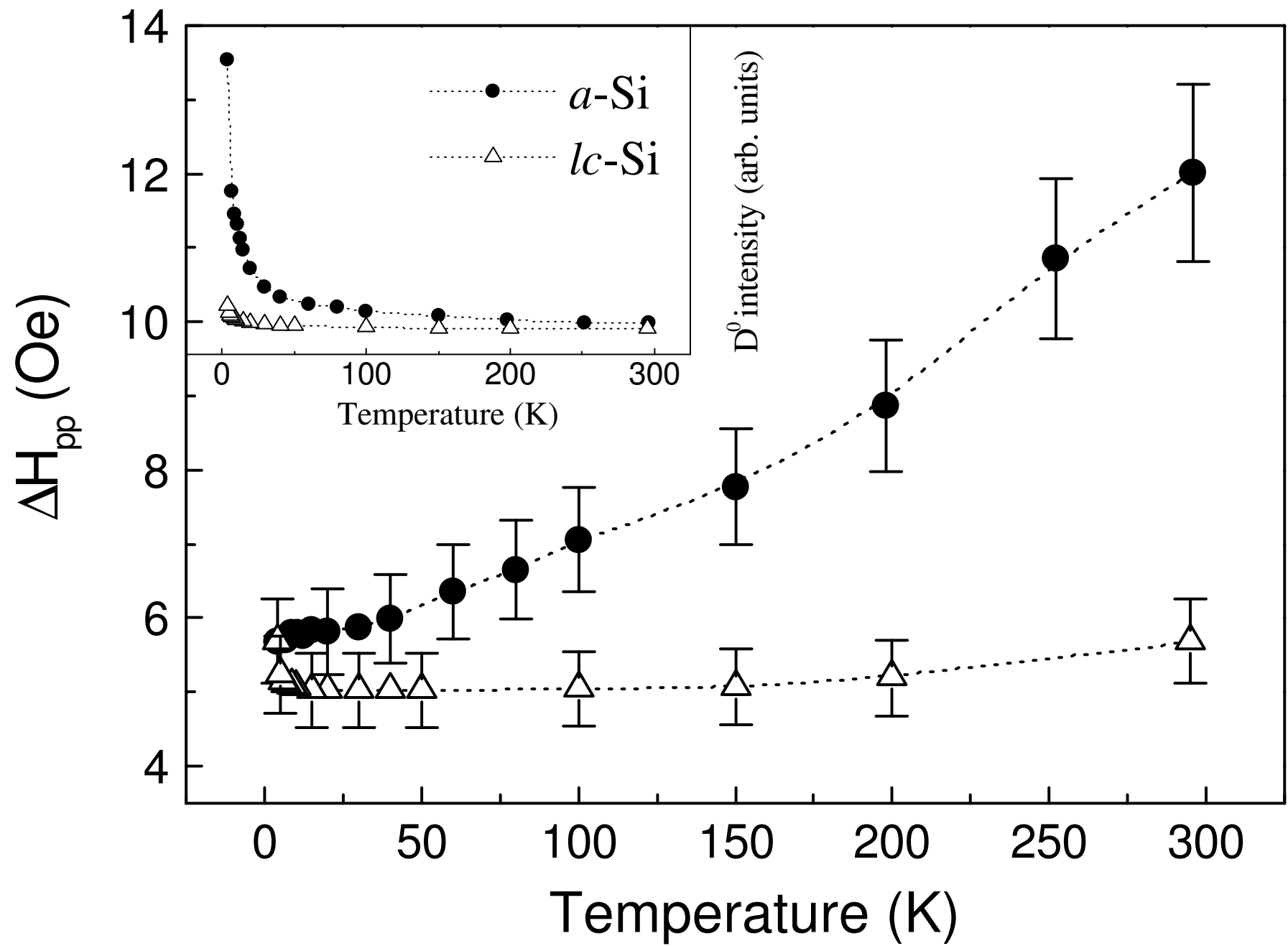
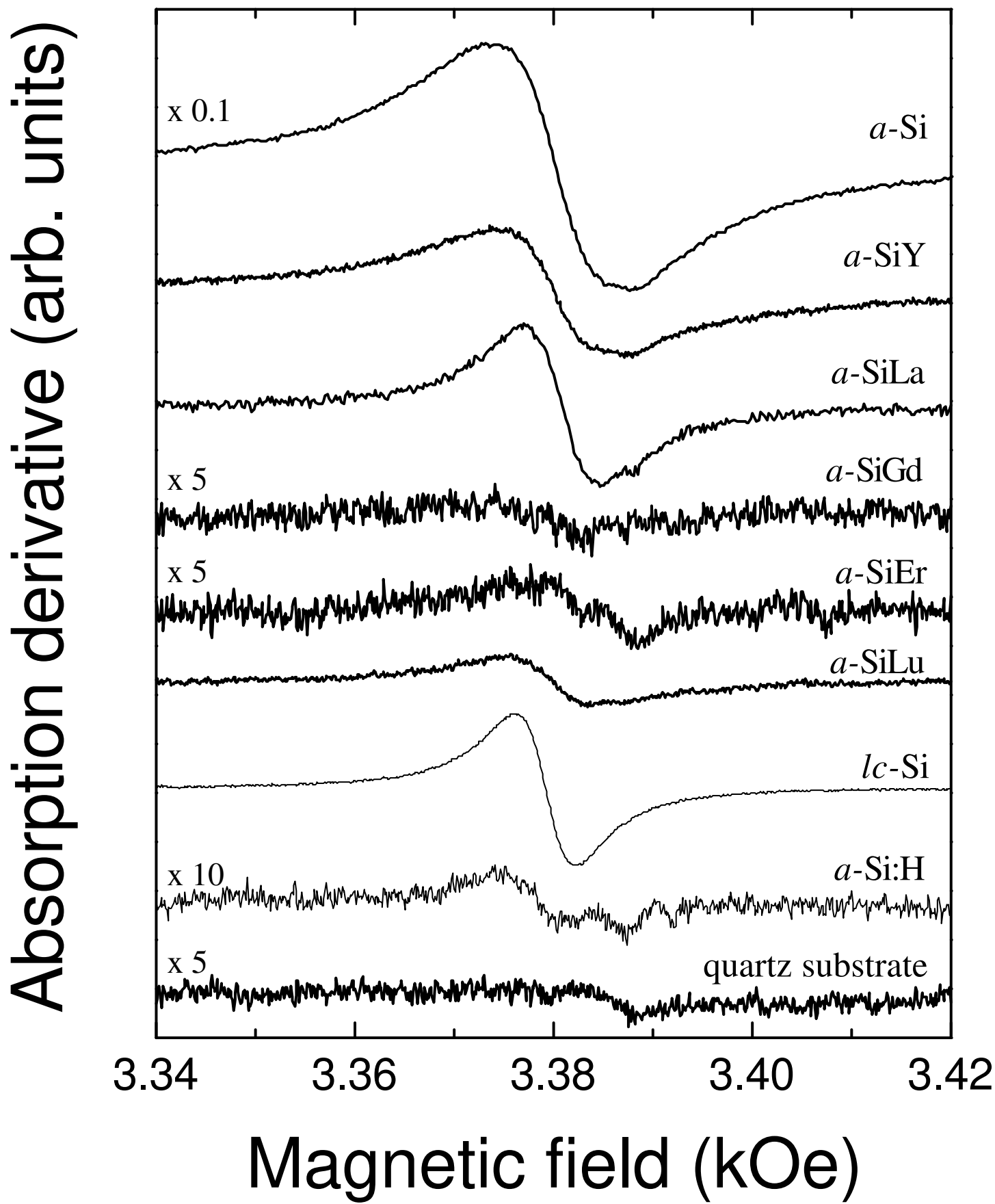
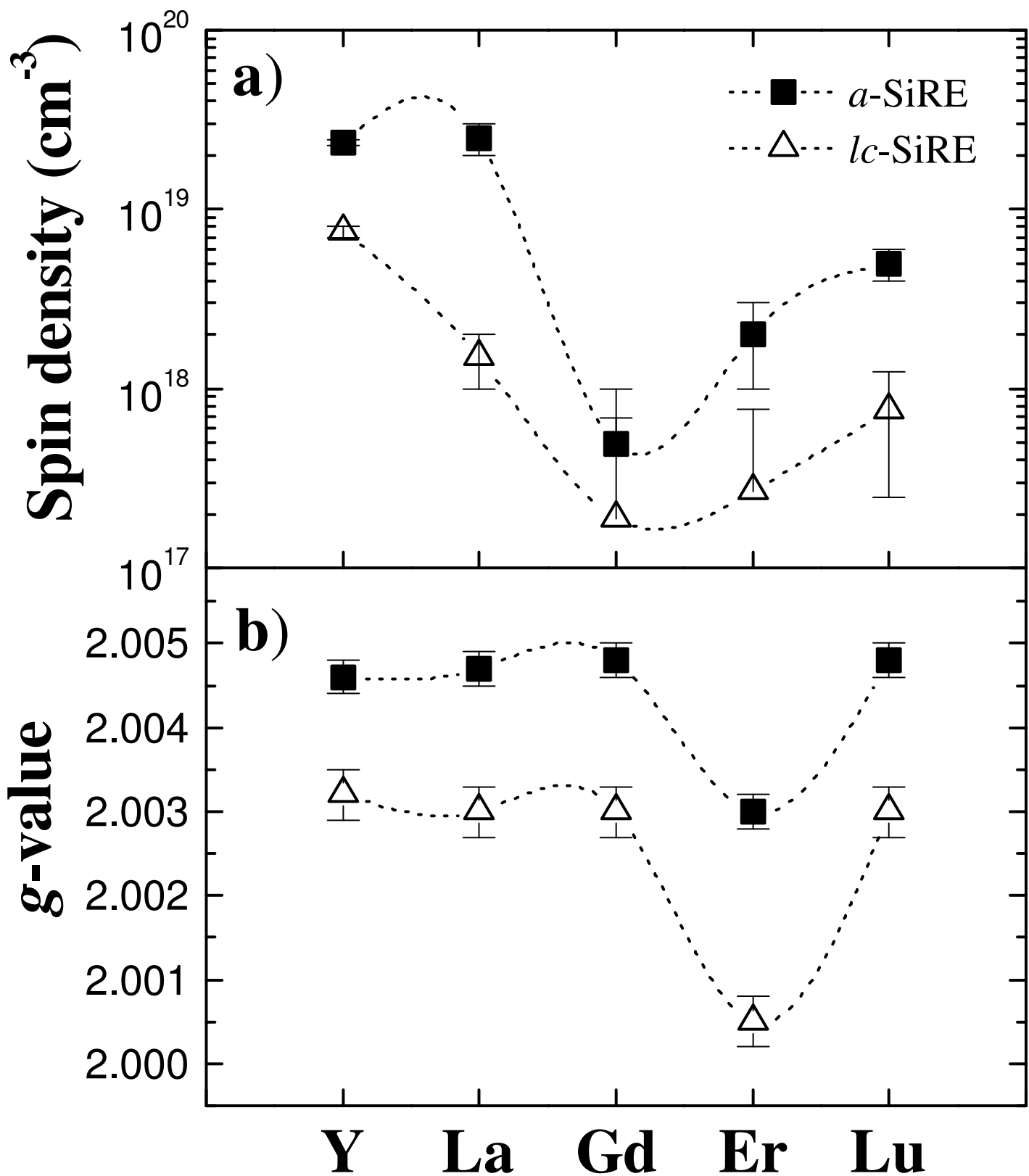
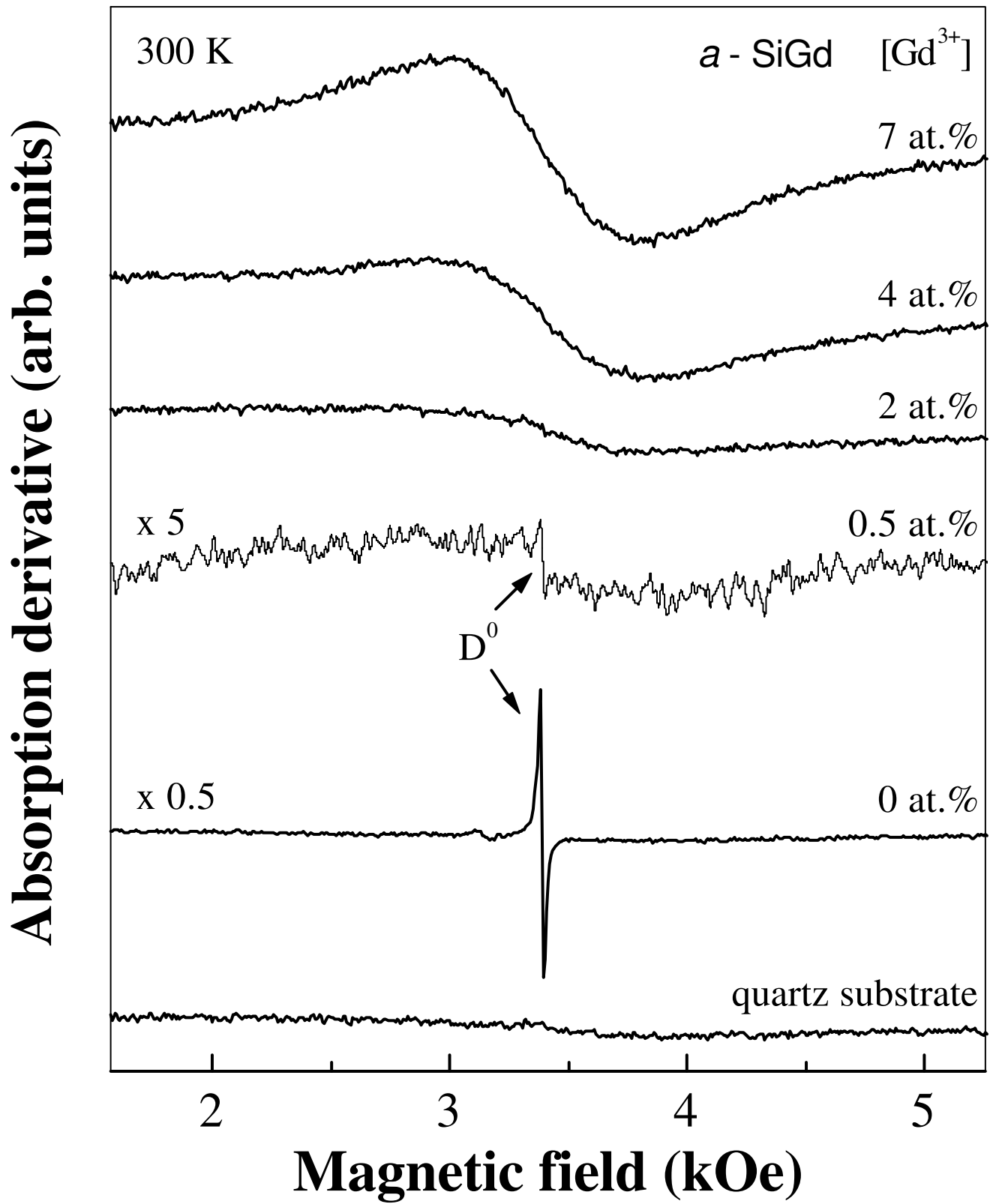
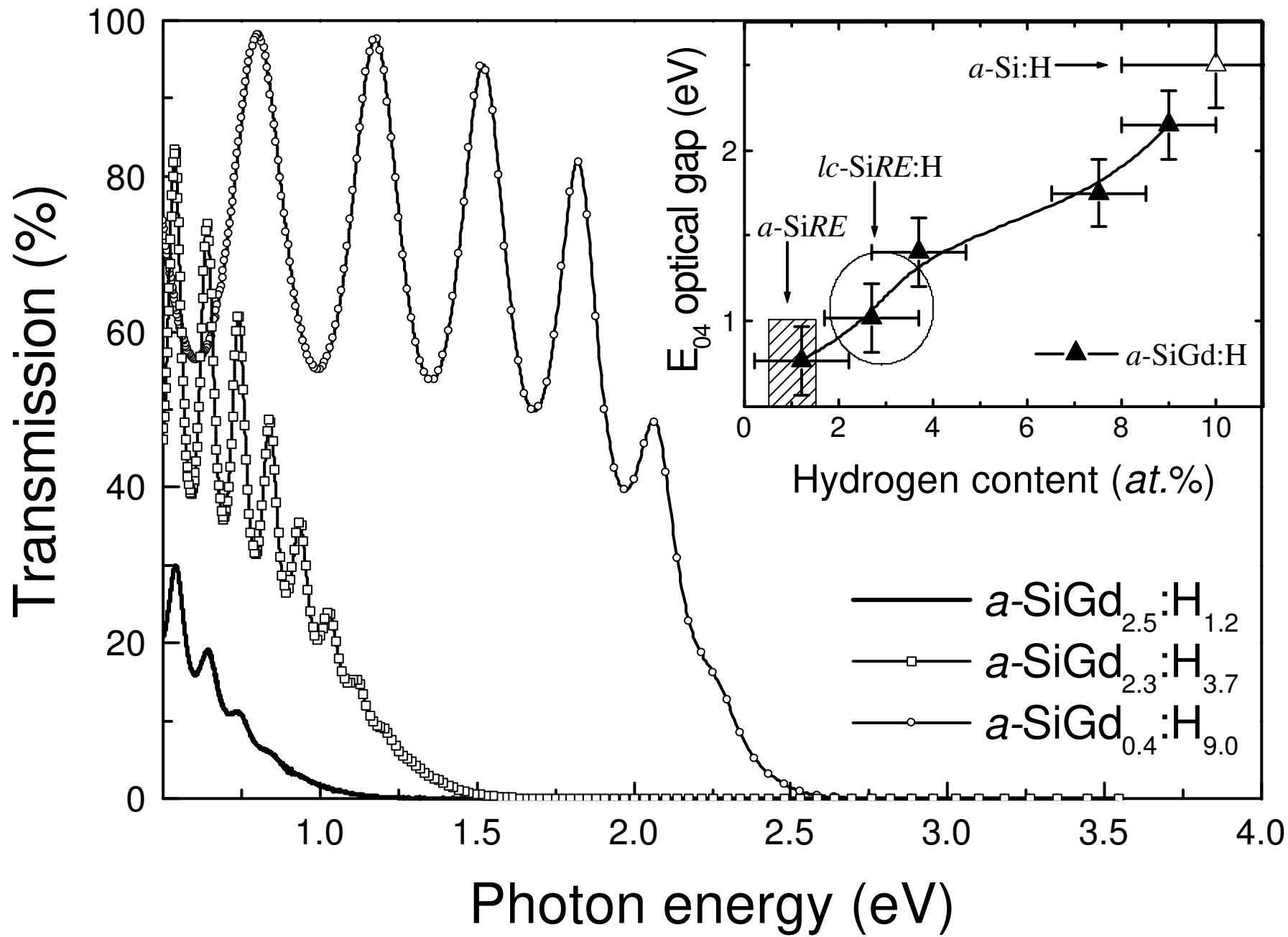
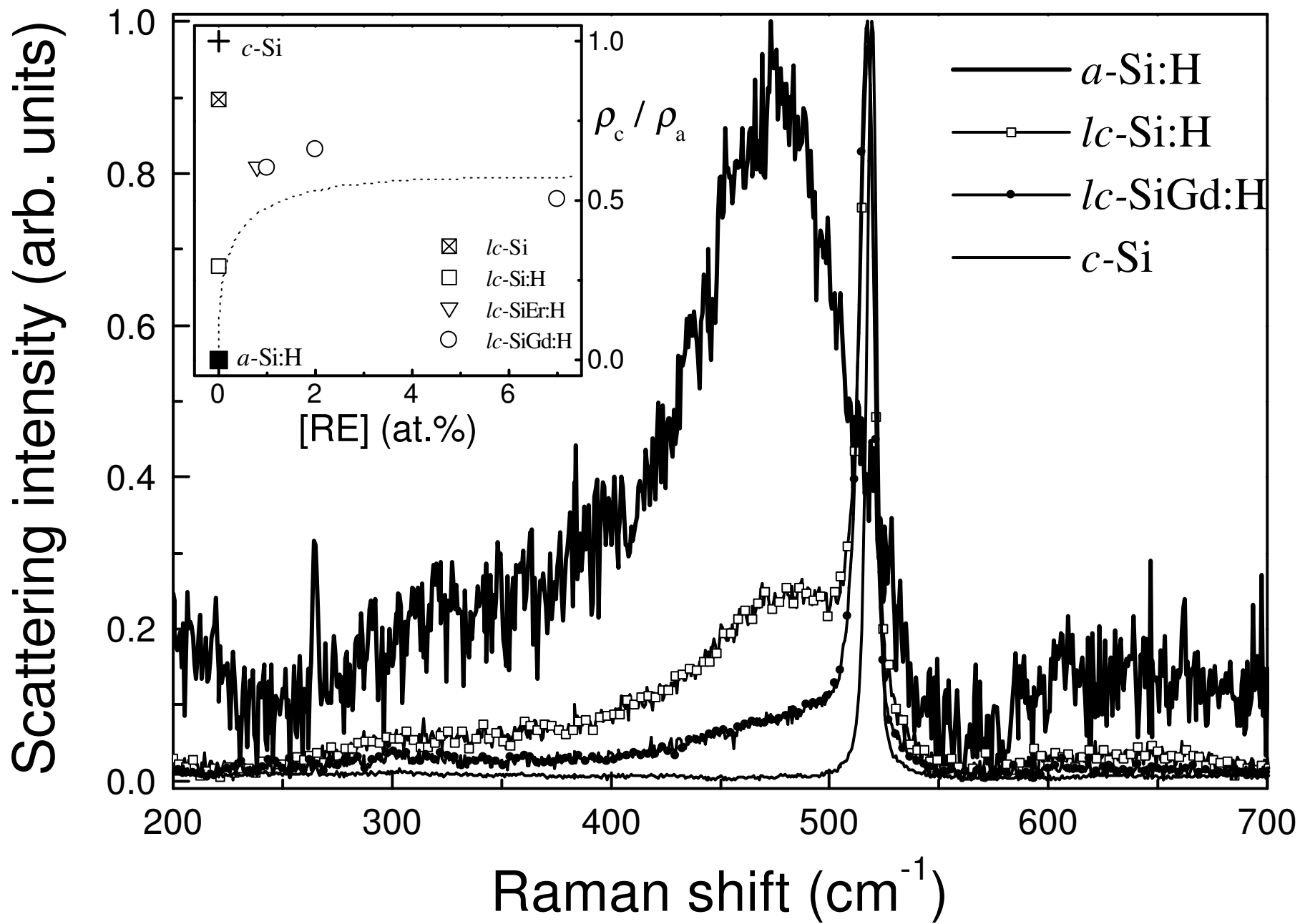
\end{document}